\documentstyle[twoside,epsfig,amssymb]{article}

\catcode`\@=11
\long\def\@makefntext#1{
\protect\noindent \hbox to 3.2pt {\hskip-.9pt  
$^{{\eightrm\@thefnmark}}$\hfil}#1\hfill}               %CAN BE USED 

\def\thefootnote{\fnsymbol{footnote}}
\def\@makefnmark{\hbox to 0pt{$^{\@thefnmark}$\hss}}    %ORIGINAL 
        
\def\ps@myheadings{\let\@mkboth\@gobbletwo
\def\@oddhead{\hbox{}
\rightmark\hfil\eightrm\thepage}   
\def\@oddfoot{}\def\@evenhead{\eightrm\thepage\hfil
\leftmark\hbox{}}\def\@evenfoot{}
\def\sectionmark##1{}\def\subsectionmark##1{}}

\oddsidemargin=\evensidemargin
\renewcommand{\thefootnote}{\fnsymbol{footnote}}

\newcounter{sectionc}\newcounter{subsectionc}\newcounter{subsubsectionc}
\renewcommand{\section}[1] {\vspace{12pt}\addtocounter{sectionc}{1} 
\setcounter{subsectionc}{0}\setcounter{subsubsectionc}{0}\noindent 
        {\tenbf\thesectionc. #1}\par\vspace{5pt}}
\renewcommand{\subsection}[1] {\vspace{12pt}\addtocounter{subsectionc}{1} 
        \setcounter{subsubsectionc}{0}\noindent 
        {\bf\thesectionc.\thesubsectionc. {\kern1pt \bfit #1}}\par\vspace{5pt}}
\renewcommand{\subsubsection}[1] {\vspace{12pt}\addtocounter{subsubsectionc}{1}
        \noindent{\tenrm\thesectionc.\thesubsectionc.\thesubsubsectionc.
        {\kern1pt \tenit #1}}\par\vspace{5pt}}
\newcommand{\nonumsection}[1] {\vspace{12pt}\noindent{\tenbf #1}
        \par\vspace{5pt}}

\newcounter{appendixc}
\newcounter{subappendixc}[appendixc]
\newcounter{subsubappendixc}[subappendixc]
\renewcommand{\thesubappendixc}{\Alph{appendixc}.\arabic{subappendixc}}
\renewcommand{\thesubsubappendixc}
        {\Alph{appendixc}.\arabic{subappendixc}.\arabic{subsubappendixc}}

\renewcommand{\appendix}[1] {\vspace{12pt}
        \refstepcounter{appendixc}
        \setcounter{figure}{0}
        \setcounter{table}{0}
        \setcounter{lemma}{0}
        \setcounter{theorem}{0}
        \setcounter{corollary}{0}
        \setcounter{definition}{0}
        \setcounter{equation}{0}
        \renewcommand{\thefigure}{\Alph{appendixc}.\arabic{figure}}
        \renewcommand{\thetable}{\Alph{appendixc}.\arabic{table}}
        \renewcommand{\theappendixc}{\Alph{appendixc}}
        \renewcommand{\thelemma}{\Alph{appendixc}.\arabic{lemma}}
        \renewcommand{\thetheorem}{\Alph{appendixc}.\arabic{theorem}}
        \renewcommand{\thedefinition}{\Alph{appendixc}.\arabic{definition}}
        \renewcommand{\thecorollary}{\Alph{appendixc}.\arabic{corollary}}
        \noindent{\tenbf Appendix \theappendixc #1}\par\vspace{5pt}}
\newcommand{\subappendix}[1] {\vspace{12pt}
        \refstepcounter{subappendixc}
        \noindent{\bf Appendix \thesubappendixc. {\kern1pt \bfit #1}}
        \par\vspace{5pt}}
\newcommand{\subsubappendix}[1] {\vspace{12pt}
        \refstepcounter{subsubappendixc}
        \noindent{\rm Appendix \thesubsubappendixc. {\kern1pt \tenit #1}}
        \par\vspace{5pt}}

\newcommand{\textlineskip}{\baselineskip=13pt}
\newcommand{\smalllineskip}{\baselineskip=10pt}

\def\eightcirc{
\begin{picture}(0,0)
\put(4.4,1.8){\circle{6.5}}
\end{picture}}
\def\eightcopyright{\eightcirc\kern2.7pt\hbox{\eightrm c}} 

\newcommand{\copyrightheading}[1]
        {\vspace*{-2.5cm}\smalllineskip{\flushleft
         }}

\def\abstracts#1#2#3{{
        \centering{\begin{minipage}{4.5in}\footnotesize\baselineskip=10pt
        \parindent=0pt #1\par 
        \parindent=15pt #2\par
        \parindent=15pt #3
        \end{minipage}}\par}} 

\def\keywords#1{{
        \centering{\begin{minipage}{4.5in}\footnotesize\baselineskip=10pt
        {\footnotesize\it Keywords}\/: #1
         \end{minipage}}\par}}

\newcommand{\bibit}{\nineit}
\newcommand{\bibbf}{\ninebf}
\renewenvironment{thebibliography}[1]
        {\frenchspacing
         \ninerm\baselineskip=11pt
         \begin{list}{\arabic{enumi}.}
        {\usecounter{enumi}\setlength{\parsep}{0pt}     
         \setlength{\leftmargin 12.7pt}{\rightmargin 0pt} %FOR 1--9 ITEMS
         \setlength{\itemsep}{0pt} \settowidth
        {\labelwidth}{#1.}\sloppy}}{\end{list}}

\newcounter{itemlistc}
\newcounter{romanlistc}
\newcounter{alphlistc}
\newcounter{arabiclistc}

\newcommand{\fcaption}[1]{
        \refstepcounter{figure}
        \setbox\@tempboxa = \hbox{\footnotesize Fig.~\thefigure. #1}
        \ifdim \wd\@tempboxa > 5in
           {\begin{center}
        \parbox{5in}{\footnotesize\smalllineskip Fig.~\thefigure. #1}
            \end{center}}
        \else
             {\begin{center}
             {\footnotesize Fig.~\thefigure. #1}
              \end{center}}
        \fi}

\newcommand{\tcaption}[1]{
        \refstepcounter{table}
        \setbox\@tempboxa = \hbox{\footnotesize Table~\thetable. #1}
        \ifdim \wd\@tempboxa > 5in
           {\begin{center}
        \parbox{5in}{\footnotesize\smalllineskip Table~\thetable. #1}
            \end{center}}
        \else
             {\begin{center}
             {\footnotesize Table~\thetable. #1}
              \end{center}}
        \fi}

\def\@citex[#1]#2{\if@filesw\immediate\write\@auxout
        {\string\citation{#2}}\fi
\def\@citea{}\@cite{\@for\@citeb:=#2\do
        {\@citea\def\@citea{,}\@ifundefined
        {b@\@citeb}{{\bf ?}\@warning
        {Citation `\@citeb' on page \thepage \space undefined}}
        {\csname b@\@citeb\endcsname}}}{#1}}

\newif\if@cghi
\def\cite{\@cghitrue\@ifnextchar [{\@tempswatrue
        \@citex}{\@tempswafalse\@citex[]}}
\def\citelow{\@cghifalse\@ifnextchar [{\@tempswatrue
        \@citex}{\@tempswafalse\@citex[]}}
\def\@cite#1#2{{$\null^{#1}$\if@tempswa\typeout
        {IJCGA warning: optional citation argument 
        ignored: `#2'} \fi}}

\def\pmb#1{\setbox0=\hbox{#1}
        \kern-.025em\copy0\kern-\wd0
        \kern.05em\copy0\kern-\wd0
        \kern-.025em\raise.0433em\box0}

\def\fnt#1#2{\footnotetext{\kern-.3em
        {$^{\mbox{\scriptsize #1}}$}{#2}}}

\def\ps@myheadings{%
    \let\@oddfoot\@empty\let\@evenfoot\@empty
    \def\@evenhead{\slshape\leftmark\hfil}%       %EVEN PAGE
    \def\@oddhead{\hfil{\slshape\rightmark}}%     %ODD PAGE
    \let\@mkboth\@gobbletwo
    \let\sectionmark\@gobble
    \let\subsectionmark\@gobble
    }
\font\tenrm=cmr10
\font\tenit=cmti10 
\font\tenbf=cmbx10
\font\bfit=cmbxti10 at 10pt
\font\ninerm=cmr9
\font\nineit=cmti9
\font\ninebf=cmbx9
\font\eightrm=cmr8

\def\qed{\hbox{${\vcenter{\vbox{                    %HOLLOW SQUARE
   \hrule height 0.4pt\hbox{\vrule width 0.4pt height 6pt
   \kern5pt\vrule width 0.4pt}\hrule height 0.4pt}}}$}}

\renewcommand{\thefootnote}{\fnsymbol{footnote}}    %USE SYMBOLIC FOOTNOTE

\def\bsc{{\sc a\kern-6.4pt\sc a\kern-6.4pt\sc a}}       %LATEX LOGO
\def\bflatex{\bf L\kern-.30em\raise.3ex\hbox{\bsc}\kern-.14em 
T\kern-.1667em\lower.7ex\hbox{E}\kern-.125em X} 

\begin{document}
\setlength{\textheight}{7.7truein}  %for 2nd page onwards

\thispagestyle{empty}

\markboth{\protect{\footnotesize\it Focusing of Opinions in the 
Deffuant-Model}}{\protect{\footnotesize\it 
Focusing of Opinions in the Deffuant-Model}}

\normalsize\textlineskip

\setcounter{page}{1}

\copyrightheading{}                     %{Vol. 0, No. 0 (1993) 000--000}

\vspace*{0.88truein}

%\fpage{1}
\centerline{\bf FOCUSING OF OPINIONS IN THE DEFFUANT MODEL:}
\vspace*{0.035truein}
\centerline{\bf FIRST IMPRESSION COUNTS}
\vspace*{0.37truein}
\centerline{\footnotesize DIRK JACOBMEIER}
\baselineskip=12pt
\centerline{\footnotesize\it Institute for Theoretical Physics, Cologne 
University}
\baselineskip=10pt
\centerline{\footnotesize\it 50923 K\"oln, Germany}
\centerline{\footnotesize\it E-mail: dj@thp.uni-koeln.de}

\vspace*{0.25truein}
\abstracts
{The paper treats opinion dynamics of an unequal distribution
as the initial opinion distribution. Simulated is the Deffuant 
model on a directed Barab\'asi-Albert network with discrete 
opinions and several subjects. Noticed is a focusing of the 
the resulting opinion distribution during the simulation towards 
the average value of the initial opinion distribution. 
A small change of the focusing is seen. 
A dependency of this change on the number of subjects and 
opinions is detected and indicates the change as a consequence
of discretization the opinions.
Hereby the average value of the initial opinion distribution
can be identified as the guide of opinion forming.
}
{}{}

\vspace*{5pt}

\setcounter{footnote}{0}
\renewcommand{\thefootnote}{\alph{footnote}}

\vspace*{1pt}\textlineskip    

\section{Introduction} 
\vspace*{-0.5pt}
\noindent
The human brain is an economically working organ. It received a lot
of informations from 'outside' (senses) and 'inside' (memories, 
associations). Not to be paralysed by working up all 
informations at the same time, it follows a strategy of stepwise
refinement. At the beginning it forms a first impression, which 
integrates more or less all informations. It goes on
in controlling and weighting all relevant informations and 
summarises them. 
At the end stands a conclusion\cite{1}.\\
People watching a movie or a performance, meeting another person, 
regarding something new, etc., do the same working method
for evaluation as the brain does. At the beginning stands a 
first impression. After leaving the theater and discussing
the movie or performance, having a talk with a new acquaintance,
examining the news more closely, they form in the end, starting from
the first impression, by checking in detail, a personal opinion.\\
Someone can take the first impression as the first truth. In discussing,
talking or examining, this first truth will be the guide (canon)
of opinion forming. 
This way of giving an opinion on truth weight, has been done by
Assmann \cite{2}, Krause and Hegselmann \cite{3}, and Malarz \cite{4}.\\
The Deffuant model offers a choice for reproducing this process.
Therefore, supposing that the impressions of all humans
are similar \cite{5}, I choose a value as the average 
of all opinions of all agents. I set this value as the first impression.\\
Therefore at the beginning of the simulation is a arrangement of opinions, 
which gives on average the chosen value of impression.

\section{Model}
\vspace*{-0.5pt}
\noindent
The model simulates a consensus forming process. The agents are connected
via a directed Barab\'asi-Albert network \cite{6} .
The opinion exchange follows Deffuant et al. \cite{7} with discrete 
opinions and several subjects (\,= questions, themes, ... ).\\
Every agent $i$ ($i\,=\,1,2,...,N$) has on each subject $S_k$ 
($k\,=\,1,2,...,S$) an opinion $O_i^k$.
The discrete opinion spectrum comprises natural numbers from 1 to O.\\
Simulations of a consensus model \'a la Deffuant on a directed
Barab\'asi-Albert network with discrete opinions have been made 
with one subject in \cite{8}, with several subjects in \cite{9,10}.

\subsection{Network assembly}
\noindent
At the beginning one knot of $m+1$ agents, each connected with all 
others, is built. 
Every newly added agent $i$ connects itself with
$m$ already existing agents in the network. The connection 
takes place stochastically. With it the probability of connecting
with an already existing agent is proportional to the total number of 
the connections of this pre-existing agent (``The rich get richer'').\\ 
Besides the connection is directed, i.e., the agents search a
partner along the $m$ connections, which they connect. The
connections, with whom they connected later when new agents are added,
can not be chosen by themselves.

\subsection{Communication}
\noindent
The communication takes place along the connections. The agents become
the active communicator $i$ in the order they have been bound into 
the network.
The partner for communication $j$ will be chosen randomly from the 
$m$ with those to whom $i$ has connected itself.
Then the over-all distance $\delta$ to the partner of communication will
be calculated. This $\delta$ results from the absolute value of the distance
of the opinions on all subjects to each other
\begin{equation}
\delta\,=\,\sum_{k=1}^S\,|O_i^k\,-\,O_j^k|\,,
\end{equation}
and is the indicator for the start of a communication:
If $\delta$ is lower or equal a given $\Delta=\,(O-1)\,S\,\varepsilon$\,
then a communication will start ($\varepsilon$ with $0\,<\,\varepsilon\,<\,1$ 
is an input parameter). Otherwise it is the next agents' turn.

\subsubsection{Rules for Simulating the Communication:} 
\noindent \label{rules}
Now agents $i$ and $j$ look randomly for a subject $S_k$ on which they 
will communicate.

\begin{itemize}
\item If the difference of opinions $(O_i^k\,-\,O_j^k)$ of both partners of
communication on the subject $k$ results in zero, then they 
agree and the communication ends.
\item If the difference of opinions equals one, one communicant will
adopt randomly the opinion from the other.
\item If the difference of opinions is larger than one, both communicants
approach each other by the amount $d$, with rounding the opinion.\\
With $d\,=\,\sqrt{1/10}\,\,\,(O_i^k\,-\,O_j^k)\,$, it will be 
$O_i^k\,:=\,O_i^k\,-\,d$ and $O_j^k\,:=\,O_j^k\,+\,d$.
\end{itemize}

\noindent
After that it is the next agents' turn.\\
The simulation ends, when during $n$ iterations over all agents no 
change of opinion in one of the communications takes place.  

\subsection{First Impression}
\noindent
'First Impression' is the initial mean opinion of all the networks agents 
opinions in all subjects $S$. The equal distribution has the median
value of the opinion-spectra $O$. To realise another than the median 
value of the opinion-spectra $O$ means to start with an unequal 
distribution of the opinions.\\
This has been done by asynchronous allocation and random displacement 
of several opinions.\\
In a second way, I choose for initializing the opinions distribution
only two possible opinions. With a probability of 50\,\% an agent gets
for all its subjects one of these two opinions. This way I call a
symmetric distribution. I have done simulations with 17 different
average network opinions generated by symmetric distributions.\\

\subsection{Parameter}
\noindent
The parameters of the model, which have been modified, are: \,\,
$\varepsilon$: tolerance,\\ $\Delta=\,(O-1)\,S\,\varepsilon$, 
$0\,\le\varepsilon\le\,1$;\,\,
$N$: Number of agents ($N$);\,\,
$S$: Number of subjects ($S$);\,\,
$O$: Number of opinions per subject ($O$),\,\,
$n$: stop criterion;\,\,the simulation stops if during $n$ consecutive 
iterations over all agents no opinion was changed.\\
The parameter of the model, which has been held constant, is \,
$m$: Number of network neighbours ($m$=3).
\begin{figure}[hb!]
\vspace*{13pt}
\centerline{\epsfig{file=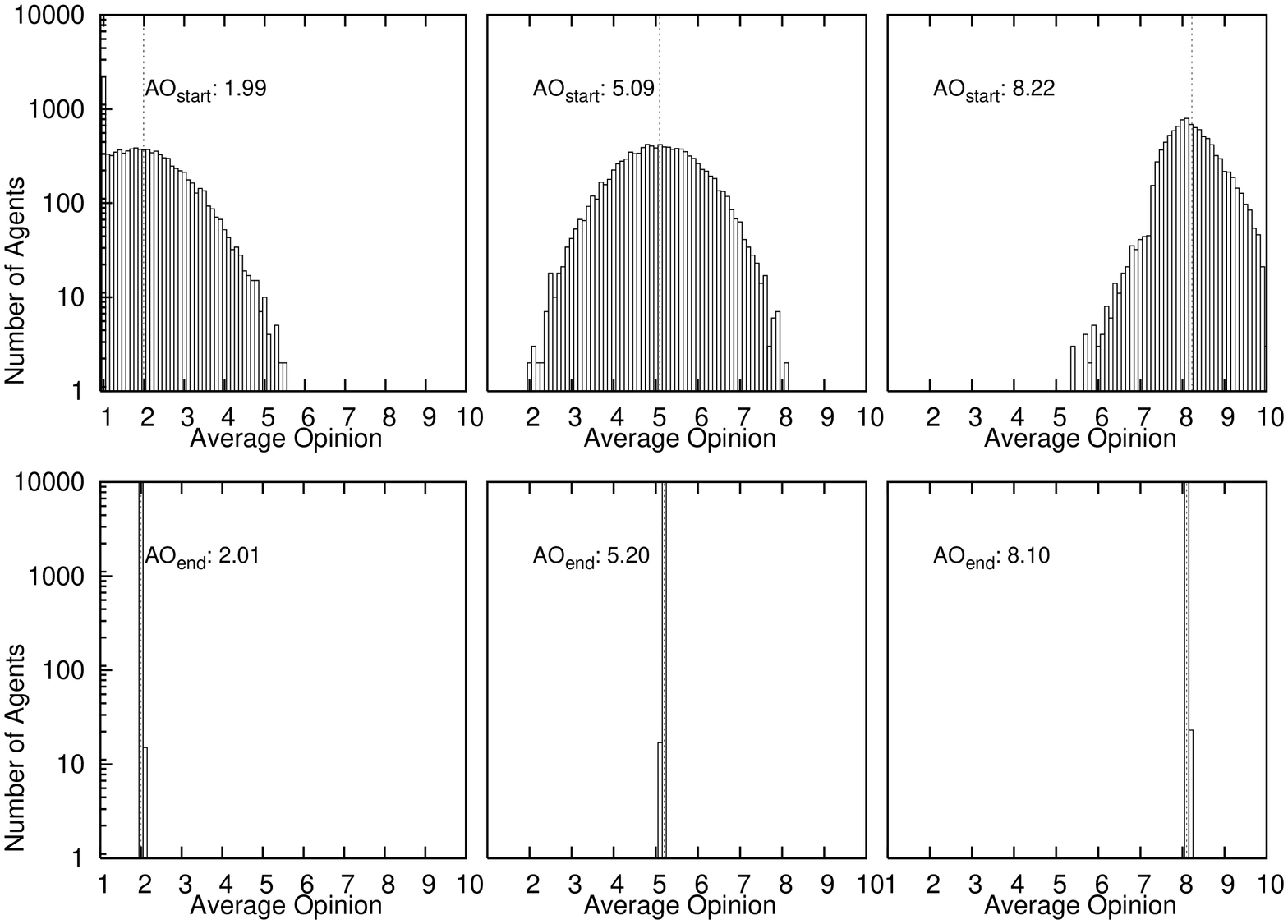,angle=-90, scale=0.52}}
\vspace*{13pt}
\fcaption{Unequal opinion distribution: Plotted are the average
opinion of the agents. Shown are also $AO_{start}$ and  $AO_{end}$.\\
The upper three graph show different opinion distributions at the
start of the simulation $AO_{start}$, the three bottom the final 
opinion distribution $AO_{end}$ of the simulation, with 
$\varepsilon$\,=\,1.0. The $y$-axis is logarithmic.
(With $N$=10003, $O$=10, $S$=10, $n$=1)
\label{VorNach}}
\end{figure}

\subsection{Methods of Evaluation}
\noindent
\begin{itemize}
\item [a)]{\bf Average Opinion $AO$}\\ The average opinion $AO$
specifies the mean of all opinions of all agents of the network 
considering all their subjects.
\begin{equation}
AO\,=\frac{1}{(S\,N)}\,\sum_{i=1}^{N}\sum_{k=1}^{S} O_i^k
\, \label{math4}
\end{equation}
The average opinion at the start of the simulation I call
$AO_{start}$, at the end of the simulation $AO_{end}$.
\item[b)]{\bf Percentage Change $PA$}\\Before the start of the
simulation, I verify the $AO_{start}$ of the network. 
After the stop of the simulation I calculate $AO_{end}$. The difference 
of $AO_{start}$ to $AO_{end}$ is given in percentage
of $AO_{end}$:
\begin{equation}
PA\,=\,100\,\left(1-\frac{AO_{start}}{AO_{end}}\right)
\, \label{math1}
\end{equation}
A positive sign implies, that the $AO_{end}$  is larger than
$AO_{start}$, a negative sign implies the reverse.
\item[c)]{\bf Standard Deviation} \\From the $PA$ \,I calculate an 
average change $\overline{PA}$ of all simulations with different 
average opinions at start.\\
Also a standard deviation of the percentage change has been 
calculated. With $iter$\,=:\,the number of simulations:
\begin{equation}
\sigma\,=\,\sqrt{\left(\frac{1}{iter-1}\right)\,{\sum_{i=1}^{iter} 
(PA_i\,-\,\overline{PA})^2 }}
\, \label{math2}
\end{equation}
\end{itemize}

\begin{figure}[ht!]
\centerline{\psfig{file=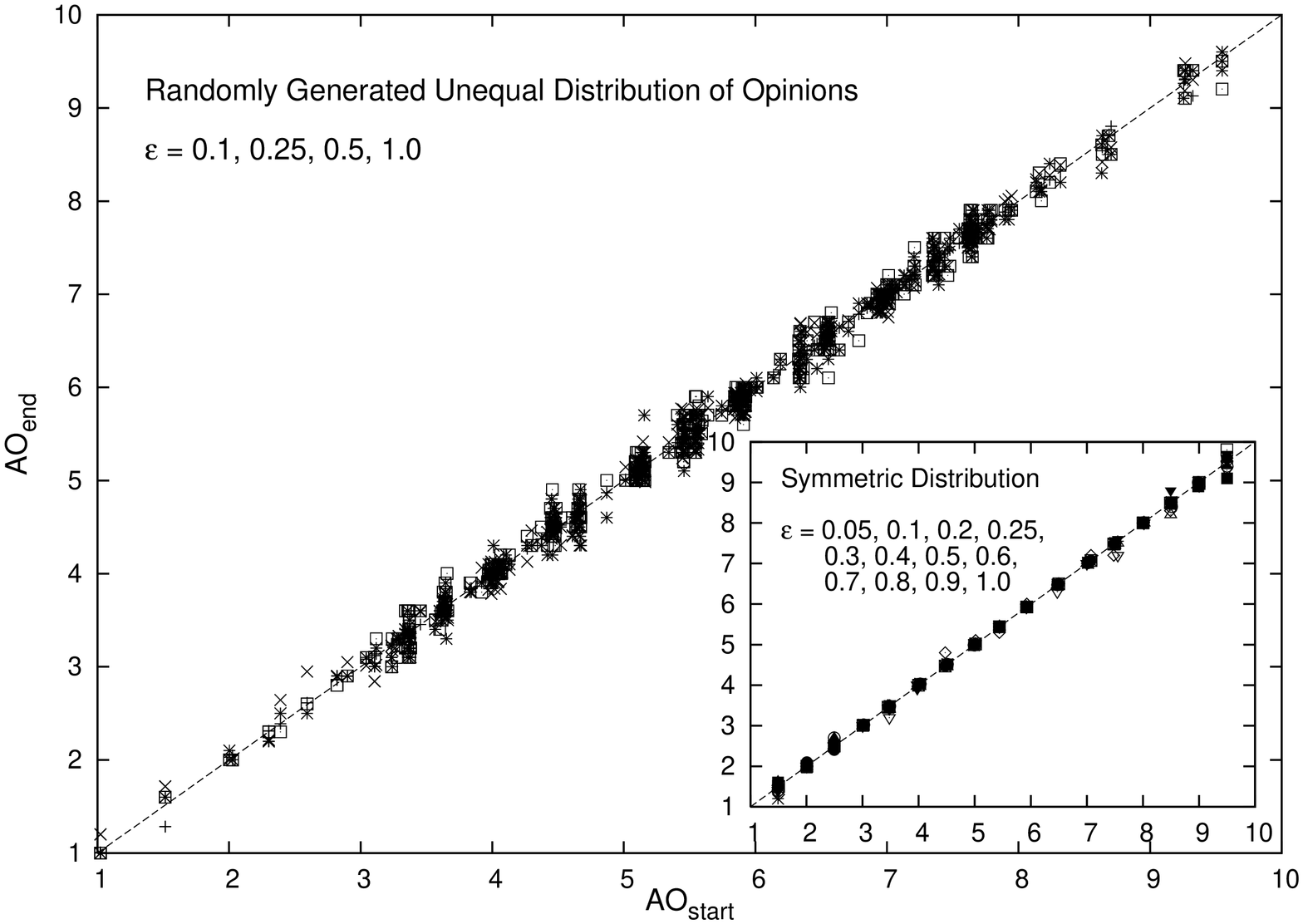, 
scale=0.52, angle=-90}}
\fcaption{The $AO_{start}$ vs. $AO_{end}$ is plotted. The tendency
of $AO_{start}$ to remain stable is obvious. The 
$AO_{end}$ varies in a small interval around the $AO_{start}$ (with
$N$=10003, $S$=10, $O$=10, $n$=1).\\
The graph shows $AO_{start}$ vs. $AO_{end}$ of an unequal opinion
distribution. In the inset is shown $AO_{start}$ vs. $AO_{end}$ of the 
simulations with the symmetric distribution.}
\label{Normal}
\end{figure}

\section{Simulation}
\vspace*{-0.5pt}

\subsection{Description}
\noindent 
The simulations have been made with $\simeq 400$ different non-equal 
opinions distributions at the start. The $AO_{end}$ of the simulations 
has been nearly the same as the $AO_{start}$ (Fig.\,\ref{VorNach} 
and \ref{Normal}).\\
The general tendency is, that $AO_{start}$ stays stable, as the 
initial distribution begin changes. 
Changing of opinions during an iteration is mostly symmetric
(see above, 2.2.1), except the second rule. Therefore with every
opinion change the mean opinion between acting agents stays
stable, except that an agent adopts randomly the opinion of the
other agent \cite{11}.   

\begin{figure}[h!]
\centerline{\psfig{file=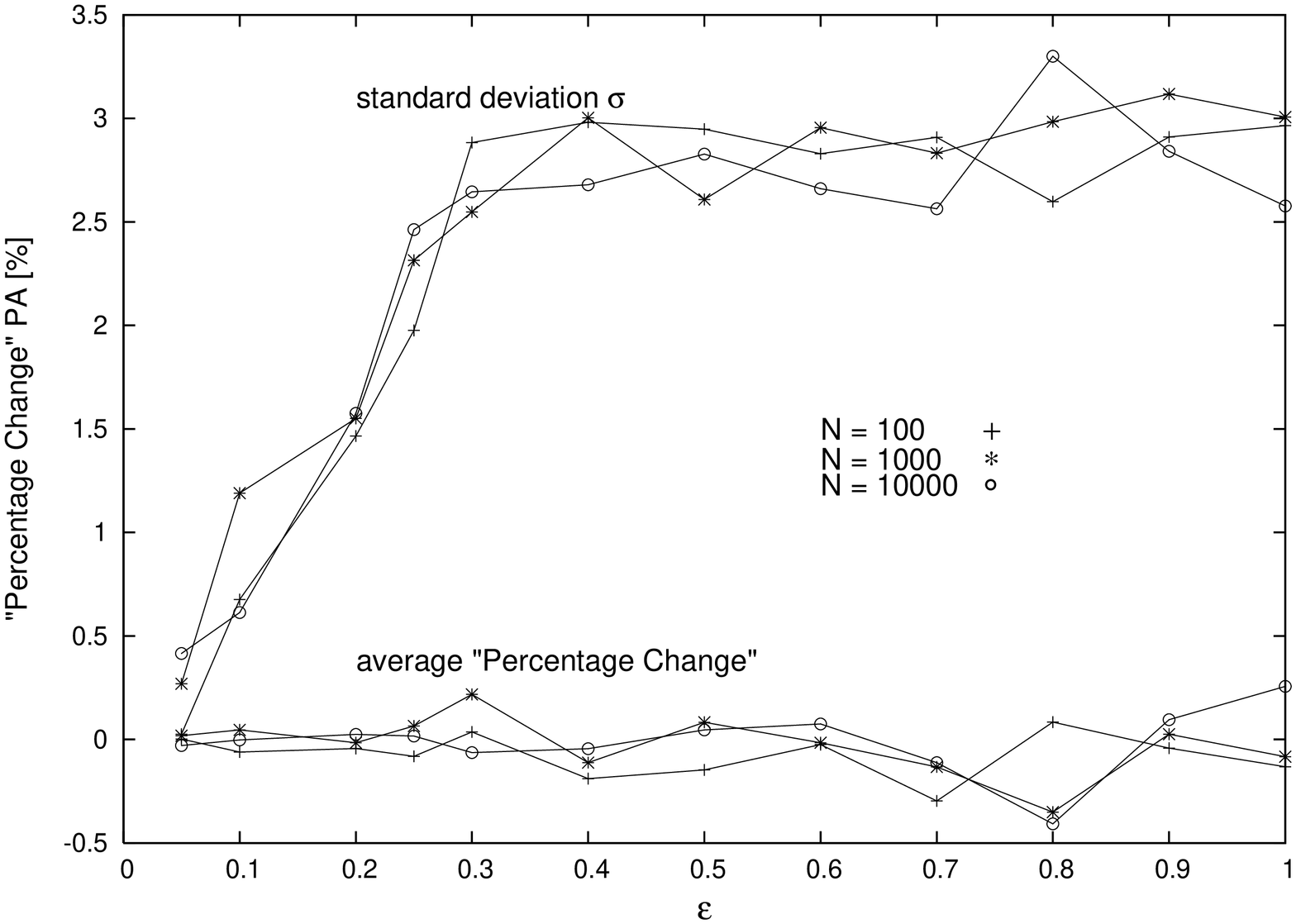, scale=0.52, angle=-90}}
\fcaption{Comparison of $\overline{PA}$ vs. $\varepsilon$ 
(curve on bottom) with the standard deviation  $\sigma$
vs. $\varepsilon$ (curve on top). The simulations have been made with 
different number of agents $N$. No obvious influence of $N$ on the
outcome is visible. (With $O$=10, $S$=10, $n$=1)}
\label{Normal_vs_N}
\end{figure}
\begin{figure}[ht!]
\centerline{\psfig{file=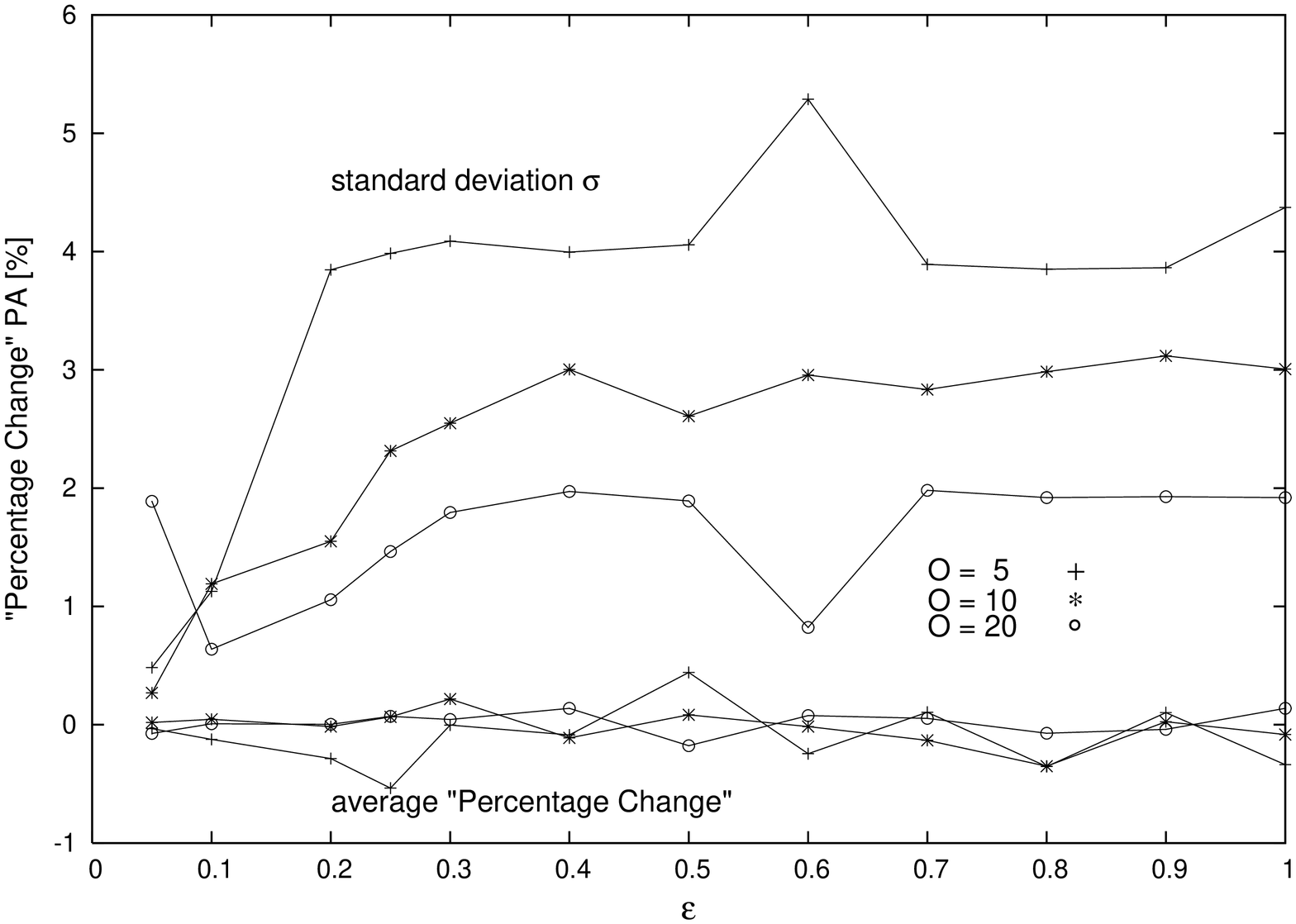, scale=0.52, angle=-90}}
\fcaption{Comparison of $\overline{PA}$ vs. $\varepsilon$ 
(curve on bottom) with the standard deviation $\sigma$ vs. $\varepsilon$ (curve
on top). The simulations have been made with different opinion-spectra
$O$. Variation of $O$ affects $\sigma$. (With $N$=1003, $S$=10, $n$=1)}
\label{Normal_vs_O}
\end{figure}
\begin{figure}[hb!]
\centerline{\psfig{file=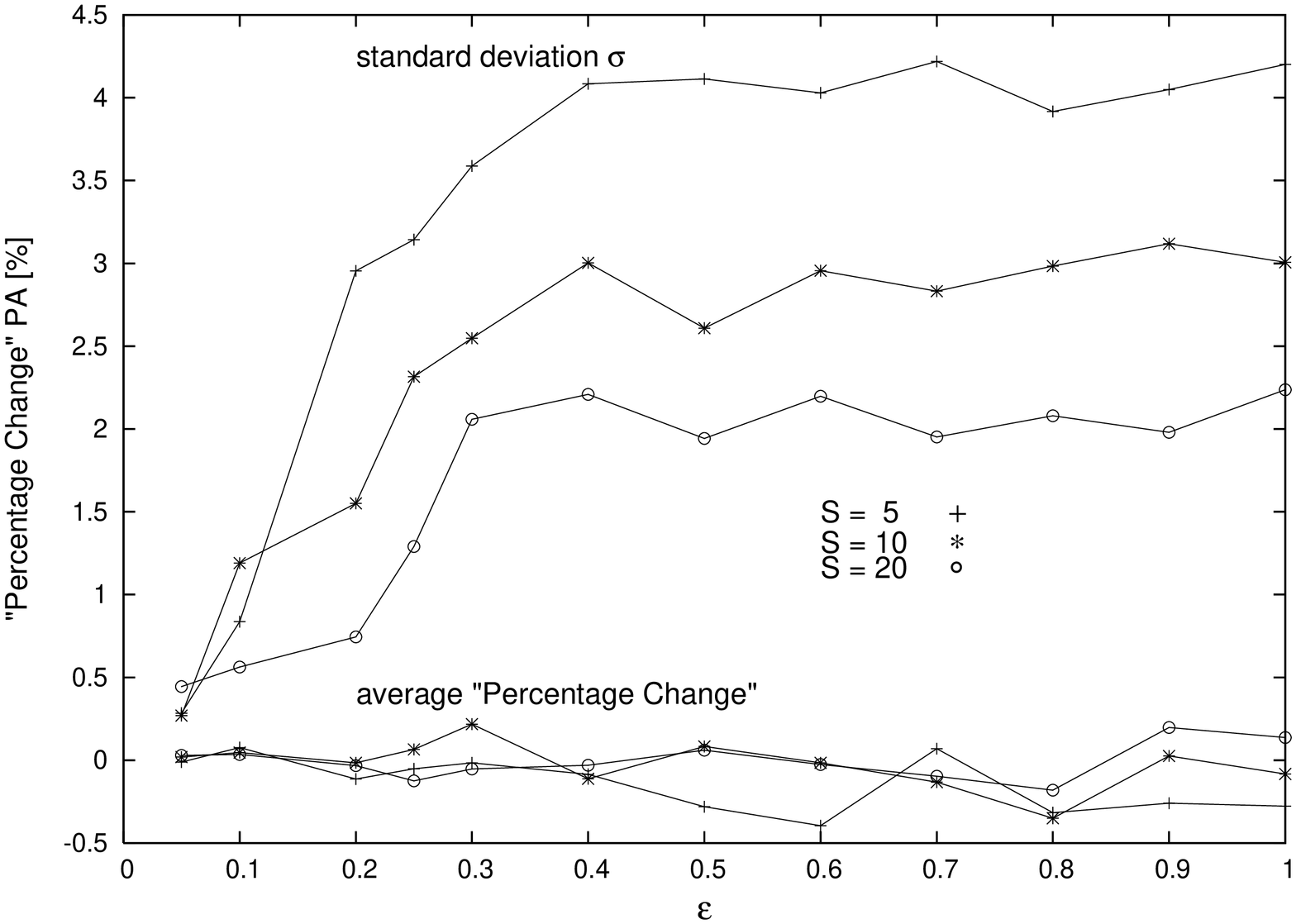, scale=0.52, angle=-90}}
\fcaption{Comparison of $\overline{PA}$ vs. $\varepsilon$ 
(curve on bottom) with the standard deviation $\sigma$ vs. $\varepsilon$ (curve
on top). The simulations have been made with different subjects
$S$. Variation of $S$ affects $\sigma$. (With $N$=1003, $O$=10, $n$=1)}
\label{Normal_vs_S}
\end{figure}

\subsection{Analysis,  Standard Deviation}
\noindent
The percentage change $PA$ has been calculated, as outlined
(Eq.\,\ref{math1}). The standard deviation (Eq.\,\ref{math2}) has been 
calculated on base of the average percentage change
$\overline{PA}$ of all simulations 
(Fig.\,\ref{Normal_vs_N} ).\\
Variations of $\varepsilon$ do influence the outcome of the 
simulations. $\overline{PA}$ stays stable around $0$, but $\sigma$
is growing with growing $\varepsilon$ until an $\varepsilon_s$
from there on $\sigma$ stays stable. $\varepsilon_s$ is identical
with the $\varepsilon$ where the minimal number of clusters of 
the network is reached and nearly all agents share the same opinions
in their subjects \cite{9}.\\
In a simulation with small $\varepsilon$ only a few opinions will
be changed, with growing $\varepsilon$ more opinions are changed.
The changing are of discrete number, this could explain the uneven
curve. 
The variation of the stop criterion $n$ and the variation of the 
number of agents $N$ (Fig.\,\ref{Normal_vs_N}) do not obviously influence 
the outcome of the simulations.
But variations of the number of subjects $S$ (Fig.\,\ref{Normal_vs_S}) 
and the opinion spectra $O$ (Fig.\,\ref{Normal_vs_O}) affect
on the outcome of the simulations.

\subsubsection{Opinion Spectra}
\noindent
$O$ divided by the smallest steps of modification of the opinions 
during the simulation gives the number of possible steps changing the
opinion. Therefore large $O$ offers more possible
values for $AO$ than small $O$, with it more possible values for
$AO_{end}$ near $AO_{start}$. This results in a smaller $\sigma$ 
with larger $O$.   

\subsubsection{Number of Subjects}
\noindent
An increasing number of $S$ results in more possible numerical values
of the average value and with it in more total values of
$AO$. Therefore large $S$ offers more possible values for $AO_{end}$ than 
smaller $S$. This also results in a smaller $\sigma$ with larger $S$.

\section{Conclusion}
\noindent
The Deffuant Algorithm is maintaining the average opinion of the
initial opinion distribution. The differences of the $AO_{start}$
to the $AO_{end}$ are small \cite{11}. We can presume the
origins of this difference in the discretization of opinions. The
difference is influenced by $S$ and $O$, due to the discretization.\\
But the most notably fact is the focus of the algorithm on the
mean value of the initial opinion distribution.
For a communicative social community means this, that the first 
impression guides the opinion forming.

\nonumsection{Acknowledgement}
\noindent
I thank greatly D. Stauffer for supporting my research in the area of 
sociophysics.

\nonumsection{References}

\end{document}